\documentstyle[12pt,psfig]{article}

\newcommand{\be}{\begin{eqnarray}}
\newcommand{\ee}{\end{eqnarray}}
\newcommand{\ep}{\varepsilon}

\newcommand{\gm}{\gamma}
\newcommand{\Gm}{\Gamma}

\newcommand{\vep}{\varepsilon}
\newcommand{\ka}{\kappa}

\newcommand{\dd}{\mbox{d}}
\newcommand{\dr}{{\rm d}}

\newcommand{\nn}{\nonumber}

\def\ppp{{\bf 3}^+}

\def\mmm{{\bf 3}^-}
\def\mmmm{{\bf 4}^-}

\begin{document}

\setcounter{page}{0}
\thispagestyle{empty}

\rightline{TTP96-32}
\rightline{hep-ph/9608407}
\rightline{August 1996}
\bigskip
\begin{center}
{\Large {\bf Threshold behavior of Feynman diagrams:}}
\end{center}
\begin{center}
{\Large{\bf the master two-loop propagator}}
\end{center}
\vspace{0.2cm}
\smallskip
\begin{center}
{\large{\sc Andrzej Czarnecki}}\footnote{E-mail:
                 {\tt ac@ttpux8.physik.uni-karlsruhe.de}}\\
{\sl Institut f\"{u}r Theoretische Teilchenphysik}\\
{\sl Universit\"{a}t Karlsruhe}\\
{\sl D--76128 Karlsruhe, Germany}\\
\vspace{0.7cm}
{\large{\sc Vladimir A. Smirnov}}\footnote{E-mail:
                {\tt smirnov@theory.npi.msu.su}}\\
{\sl Nuclear Physics Institute}\\
{\sl Moscow State University}\\
{\sl Moscow 119899, Russia}\\
\vspace{0.8cm}
{\large{\bf Abstract}}\\
\end{center}
\vspace{0.1cm}
An asymptotic expansion of the two-loop two-point ``master'' diagram
with two masses $m$ and $M$, on the mass shell $Q^2=M^2$, is
presented.  The treatment of the non-analytical terms
arising in the expansion around the branching point  is discussed.
Some details of
the calculation of a new class of two-loop integrals are given.
\newpage

\section{Introduction}\label{intro} Phenomenological problems in
particle physics often require calculations of multi-loop
Feynman diagrams involving fields of various masses. Since
exact results for diagrams with more than one mass scale
practically do not exist beyond one loop, it is reasonable to
apply some analytical approximations. In particular, one
successfully applies explicit formulae for asymptotic expansions
in various 
limits of momenta and masses \cite{ae} (see \cite{vs-review}
for an informal review).  The underlying idea is to utilize a
hierarchy of mass scales in the given problem to reduce it to a
calculation of one-scale diagrams.  There are two general
classes of problems which can be solved using those expansions:
the large momentum case (when some external momenta are much
larger than other relevant mass scales) and the case of very
heavy virtual particles. In both classes important phenomenological
problems have recently been solved by asymptotic expansions
\cite{appl1}. See also \cite{appl2} where propagator and vertex
two-loop diagrams were systematically calculated in various regions
of momenta and masses.

There is, however, an important class of diagrams with a heavy
particle both inside and on its mass shell on the external legs, when
off-shell formulae \cite{ae} are generally not
applicable\footnote{Although these formulae are guaranteed at least
off the mass shell, they are also valid in some on-shell situations,
for example, in the pure large mass limit.}. Explicit formulae for
asymptotic expansions in momenta and masses in some typical limits
with large on-shell momenta were presented in \cite{vs}.  In
our paper we discuss the two-point on-shell integrals with two
mass scales shown in Fig.~1 ($m<M$).

\vspace*{0mm}
\begin{figure}[htb]
\hspace*{-5mm}
\begin{minipage}{16.cm}
\[
\mbox{
\hspace*{10mm}
\begin{tabular}{cc}
\hspace*{10mm}
\psfig{figure=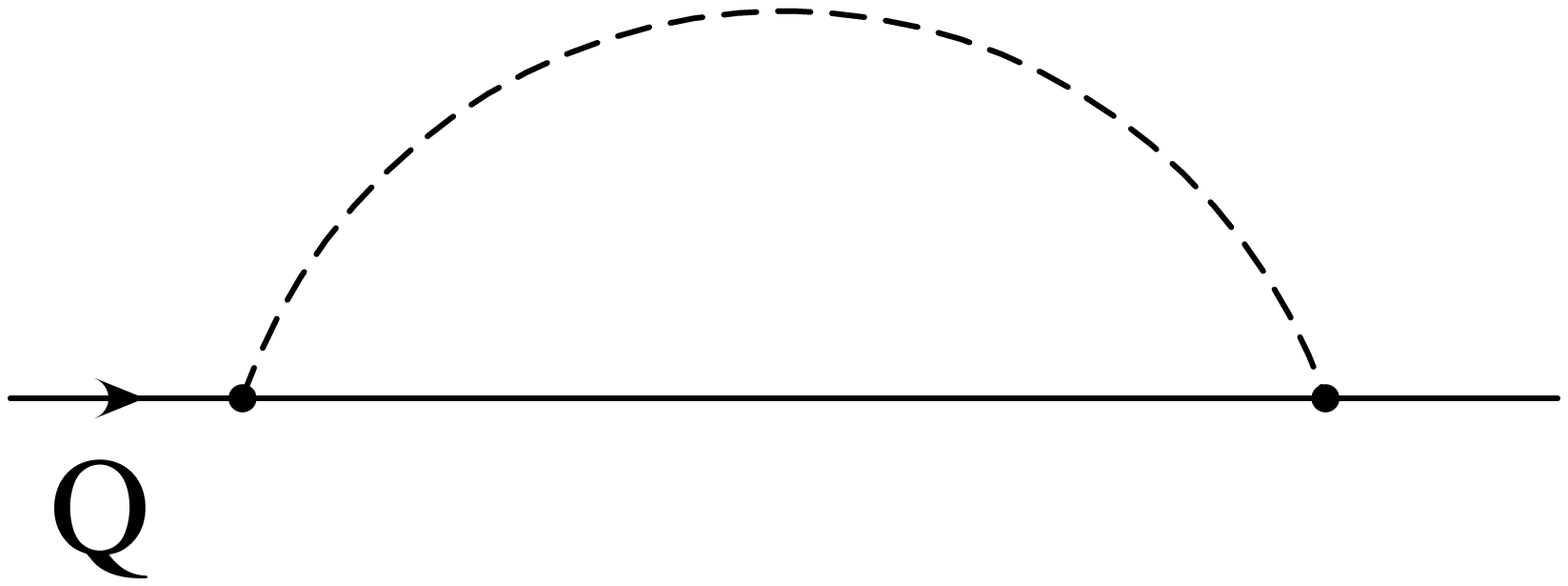,width=40mm,bbllx=210pt,bblly=410pt,%
bburx=630pt,bbury=550pt} 
&\hspace*{25mm}
\psfig{figure=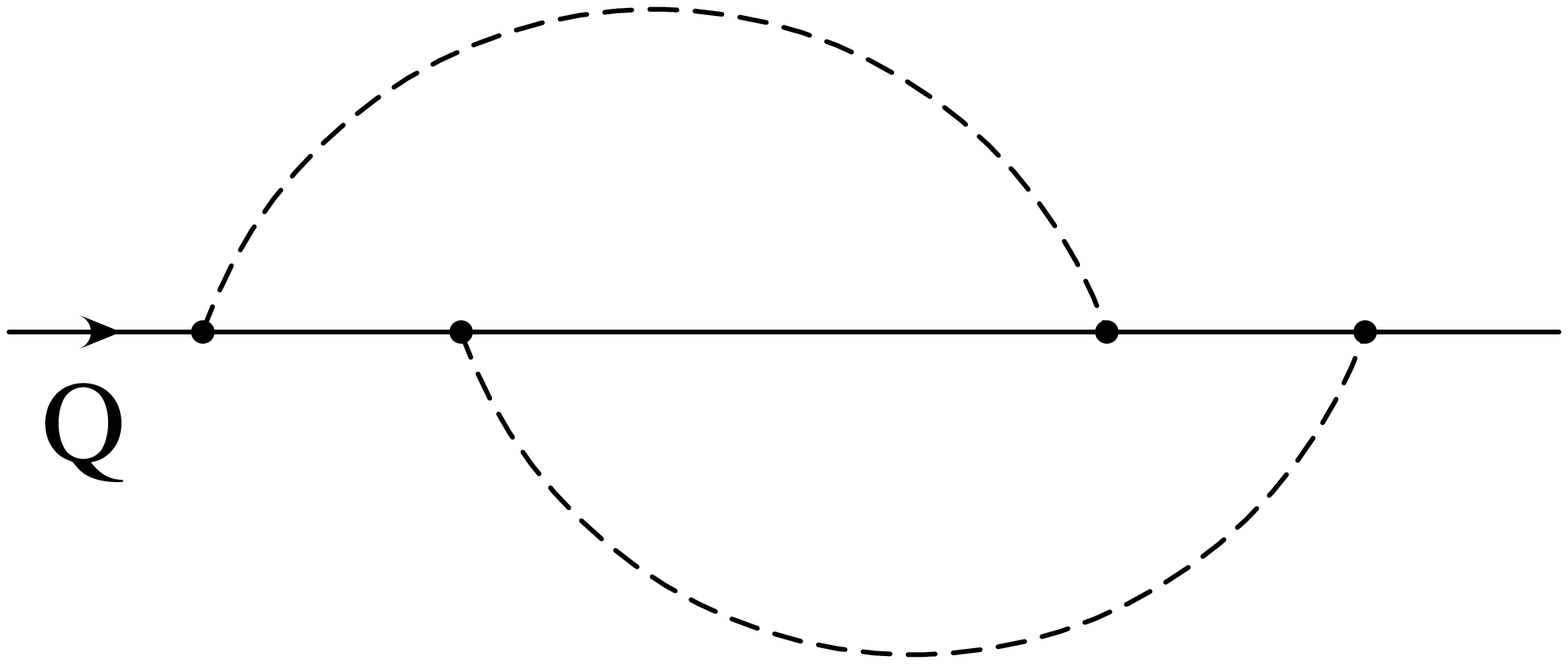,width=40mm,bbllx=210pt,bblly=410pt,%
bburx=630pt,bbury=550pt}
\\[15mm]
\hspace*{-12mm}
(a) & (b)
\end{tabular}}
\]
\end{minipage}
\caption{Examples of two-point on-shell ($Q^2=M^2$) 
  diagrams with two mass scales;
  solid lines denote a heavy particle with mass $M$, and dashed lines
  -- a light one with mass $m$.} 
\label{figs}
\end{figure}

The one-loop diagram in Fig.~1a is a good example which illustrates
both the difficulties of the previous approaches and the idea of the
new method.  Of course the exact result is well known:
\be
G_1 &=& \int {\dr^D k \over (k^2-m^2)[(k+Q)^2-M^2]}
\nonumber\\
&=&
{i\pi^{D/2}} \left(
{1\over \ep}+2-\ln M^2 -x^2\ln x -2x\sqrt{4-x^2}\arctan
\sqrt{{2-x\over 2+x}}\right)
\ee
with $x=m/M$. In this paper
we use dimensional regularization with $D=4-2\ep$; it is understood
that the masses in propagators have an infinitesimal negative
imaginary part which is not displayed.  We drop Euler's $\gamma$ in the
formulae.

Suppose now we want to find the expansion of $G_1$ around $x=0$ without
calculating the above two-scale integral.  Unfortunately it is
impossible to apply just the Taylor expansion of the {\em
  integrand} around $m=0$; the expanded propagator ${1\over k^2-m^2}=
{1\over k^2}\sum_{n=0}^\infty \left( {m^2\over k^2}\right)^n$ leads to
infrared divergences.  This difficulty is connected with the
impossibility of the naive change of order of expansion and integration.
General formulae of asymptotic expansions give, for this diagram, {\em two}
contributions: besides this `naive' Taylor expansion in $m$ one has
a contribution which is the Taylor expansion of another factor of
the integrand, ${1\over k^2+2kQ}$, around $k^2=0$ \cite{vs}.
(In the off-shell situation, this extra contribution would be \cite{ae}
just the Taylor expansion of ${1\over (k+Q)^2-M^2}$ around $k=0$,
which happens to be impossible now because $Q$ is {\em
 on-shell}, $Q^2=M^2$.)

We therefore expand ${1\over k^2+2kQ}$ around $k^2=0$
and obtain ${1\over k^2+2kQ}={1\over 2kQ}\sum_{n=0}^\infty
\left(- {k^2\over
  2kQ}\right)^n$.  The resulting integrals are homogeneous in $Q$
and depend on one mass scale $m$ only.  It is easy to convince
oneself that in the result the divergences generated in the Taylor
expansions are canceled and the sum of both contributions leads to
\be
G_1 =
{i\pi^{D/2}} \left[
 {1\over \ep}+2-\ln M^2 -\pi x + x^2(1-\ln x) +{\pi\over 8}x^3 +{\cal
   O}(x^4)
\right] \, .
\label{eq:1loop}
\ee
We notice that the terms $\pi \left({m\over M}\right)^{2n+1}$ appear
even though $m$ is found in the integrand  only in an even power.
These non-analytical terms arise because we expand the diagram around
the point $m=0$ which is also a branching point (threshold) of this
Feynman diagram.

For the two-loop diagram in Fig.~1b no exact formula is known.  In
this paper we give its asymptotic expansion in the limit $m/M\to
0$. The next section is rather formal; it describes the general
formalism of the expansion.  In section \ref{sec:calc} this machinery
is applied to the diagram 1b; it turns out that there are four
contributions (rather than two, as in the one-loop case).  The
calculation of a new type of integrals which arises in this context is
discussed in some detail.

\section{Explicit formula of asymptotic expansion}

We consider the Feynman integral $F_{\Gm}$ corresponding to a graph
$\Gm$ when the masses $M_i$ and external momenta $Q_i$ are large
with the respect to small masses $m_i$ and external momenta $q_i$.
Let us suppose that the momenta are non-exceptional.
Let the large external momenta be on the mass shell, $Q_i^2=M_i^2$.
Then the asymptotic expansion in the limit $Q_i,M_i\to \infty$ takes
the following explicit form \cite{vs}:
\be
F_{\Gm} (Q_i, M_i, q_i, m_i;\ep)
\; \stackrel{\mbox{\footnotesize$M_i \to \infty$}}{\mbox{\Large$\sim$}} \;
\sum_{\gamma}  {\cal M}_{\gm}
F_{\Gm} (Q_i, M_i, q_i, m_i;\ep).
\label{eae}
\ee
Here the sum is over subgraphs $\gm$ of $\Gm$ such that

(a) in $\gm$ there is a path between any pair of external vertices
associated with the large external momenta $Q_i$;

(b) $\gm$ contains all the lines with the large masses;

(c) every connectivity component $\gm_j$ of
the graph $\hat{\gm}$ obtained from $\gamma$ by collapsing
all the external vertices with the large external momenta to a point
is 1PI with respect to the lines with the small masses.

In general $\gm$ can be disconnected. One can distinguish
the connectivity component $\gm_0$ which contains the external
vertices with the large momenta.

The operator ${\cal M}_{\gm}$ that is involved in the sum is a product
$\prod_i {\cal M}_{\gm_i}$ of operators of Taylor expansion in
certain momenta and masses. These operators are by definition applied
to integrands of Feynman integrals over loop momenta.
For connectivity components $\gm_i$ other than $\gm_0$ the
corresponding operator performs Taylor expansion of the Feynman integral
$F_{\gm_i}$ in its small masses and external momenta. (Note that its small
external momenta are generally not only the small external momenta of the
original Feynman integral but also some loop momenta of $\Gm$.)
Consider now ${\cal M}_{\gm_0}$. The component $\gm_0$ can be naturally
represented as a union of its 1PI components and cut lines
(after a cut line is removed the subgraph becomes disconnected;
here they are of course lines with the large masses). By definition
${\cal M}_{\gm_0}$ is again factorized and the Taylor expansion of
the 1PI components of $\gm_0$ is performed as in the case of other
connectivity components
$\gm_i$.

It suffices now to describe the action of the
operator ${\cal M}$ on the cut lines. Let $l$ be such a line,
with a large mass $M_i$, and let its momentum be $P_l+k_l$ where
$P_l$ is a linear combination of the large external momenta and
$k_l$ is a linear combination of the loop momenta and small external momenta.
If $P_l=Q_i$ then the operator $\cal M$ for this component of $\gm$ is
\be
\left. {\cal T}_{\ka} \frac{1}{\ka k_l^2+2Q_i k_l} \right|_{\ka=1} \, .
\ee
By ${\cal T}_x$ we denote the operator of the Taylor expansion
in $x$ around $x=0$.
In all other cases, e.g. when $P_l =0$, or it is a sum of two or more
external momenta, the operator ${\cal M}$ reduces
to the ordinary Taylor expansion in small (with  respect to this line
considered as a subgraph) external momenta, i.e.
\be
\left. {\cal T}_{k} \frac{1}{ ( k_l+P_l)^2-M_i^2} \equiv
{\cal T}_{\ka} \frac{1}{ (\ka k_l+P_l)^2-M_i^2} \right|_{\ka=1} \, .
\ee

In all cases apart from the cut lines with $P_l^2 =M^2_i$
the action of the corresponding operator $\cal M$ is graphically
described by  contraction of the corresponding subgraph to a point and
insertion of the resulting polynomial into the reduced vertex of the
reduced graph.

\section{The two-loop master diagram}
\label{sec:calc}

Let $\Gm$ be the two-loop self-energy graph (see Fig.~1b) containing
three heavy lines with the mass $M$, two light lines with the mass $m$ and
let the external momentum be on the mass shell, $Q^2=M^2$:
\be
G_2=
\int\!\! \int \frac{\dd^D k \dd^D l}{(k^2-m^2)(l^2-m^2)
(k^2+2Qk)
\left[(k+l)^2 + 2Q(k+l)\right]
(l^2+2Ql)} .
\label{FI2}
\ee

According to the general formula (\ref{eae}) three types of
subgraphs contribute to the asymptotic expansion in the limit $m/M\to 0$:

(i) the graph $\Gm$ itself;

(ii) the subgraph $\gm_1$ consisting of the left
triangle and the third heavy line as well as symmetric
subgraph $\gm_2$;

(iii) the subgraph $\gm_3$ consisting of three heavy lines.

For (i) we expand  the integrand around $m=0$:
\be
\lefteqn{
\int \!\! \int
\frac{\dd^D k \dd^D l}{(k^2+2Qk)((k+l)^2+2Q(k+l))(l^2+2Ql)}
{\cal T}_{m^2} \frac{1}{(k^2-m^2)(l^2-m^2)}
}
\nn \\
&&\hspace*{80mm}
\equiv \sum_{n=0}^{\infty} m^{2n} \sum_{j=0}^{n} I(j,n-j)\, .
\label{ci}
\ee
where
\be
I(a,b) = \int \!\! \int \frac{\dd^D k \dd^D l}
{(k^2)^{a+1}
(l^2)^{b+1}
(k^2+2Qk)
\left[(k+l)^2+2Q(k+l)\right]
(l^2+2Ql)} .
\label{Jab}
\ee
The leading term (which gives $G_2$ at $m=0$) is equal to \cite{DB}
\be
I(0,0) = -{\pi^D\over M^2}
\left[ \frac{3}{2} \zeta(3) - \pi^2 \ln 2 \right]
  +{\cal O}(\ep).
\label{J00}
\ee
For any $a$ and $b$ the integral $I(a,b)$ can be reduced to
$I(0,0)$ and one-loop integrals  using recurrence relations
\cite{IBP,bro91a}.  This algorithm has been implemented
\cite{b2c} in FORM \cite{form}.

Each of the two contributions to (ii) equals
\be
\lefteqn{\int \frac{\dd^D l}{l^2-m^2}
{\cal T}_{l^2} \frac{1}{l^2+2Ql}
\int \frac{\dd^D k }{k^2+2Qk}
{\cal T}_{m^2} \frac{1}{k^2-m^2}
{\cal T}_{l}\frac{1}{(k+l)^2+2Q(k+l)}}
\label{cii}
\\
&=&
\sum_{j_1,j_2,j_3=0}^{\infty}
(-1)^{j_1}  m^{2(j_1+j_2)}
\sum_{n=0}^{j_3}
{j_3 \choose n}
\nonumber\\
&&\qquad \qquad \times
\int \dd^D l \frac{(2Ql+m^2)^{j_3-n}}
{(l^2-m^2)(2Ql)^{j_1+1}}
\int \dd^D k \frac{(2kl)^n}{(k^2)^{j_2+1}(k^2+2Qk)^{j_3+2}} .
\label{ciiexp}
\ee
The factors $l^2$ in the numerator have been
substituted by $m^2$ because whenever
$l^2-m^2$ appears it cancels the corresponding factor in the denominator
and we get a massless tadpole integral which vanishes.  The
calculation of both one-loop integrals in (\ref{ciiexp}) poses no
difficulties.

Finally, the contribution (iii) is
\be
\int\!\! \int \frac{\dd^D k \dd^D l}{(k^2-m^2)(l^2-m^2)} \left.
{\cal T}_{\ka}
\frac{1}{(\ka k^2+2Qk)(\ka (k+l)^2+2Q(k+l))(\ka l^2+2Ql)}
\right|_{\ka=1}
\nn \\
\equiv \sum_{j_1,j_2,j_3=0}^{\infty}
\int \!\! \int \dd^D k \dd^D l \frac{(-m^2)^{j_1+j_2}
\left[-(k+l)^2\right]^{j_3}}
{(k^2-m^2)
(2Qk)^{j_1+1}
(2Q(k+l))^{j_3+1}
(l^2-m^2)
(2Ql)^{j_2+1}} \, .
\label{iii}
\ee

In this formula the following new type of integrals arises:
\be
\lefteqn{J(a_1,a_2,a_3,a_4,a_5) = }
\nonumber\\ &&
\int\!\!\int
\dr^Dk\dr^Dl
{(kl)^{a_5}
\over
(k^2-m^2+i0)^{a_1}
(l^2-m^2+i0)^{a_2}
(2Qk+i0)^{a_3}
(2Qk+2Ql+i0)^{a_4}} \, .
  \label{eq:defiii}
\ee
In general we can also have a power of $2Ql$ as an extra term in the
denominator but this can be removed by partial fraction decomposition.
Any such integral can be calculated analytically.
We are interested in the case $a_1=a_2=1$.
In calculating (\ref{eq:defiii}) we first reduce the problem to the case
$a_5=0$ by expressing the product $(kl)^{a_5}$ in terms of traceless
products $(kl)^{(i)}\equiv k^{(\alpha,i)} l_{(\alpha,i)}$.
Then we notice that the result of the integration over $l$ of a
traceless product $l_{(\alpha,i)}$ times the part of
the integrand which depends only on $l$ must be proportional to the
traceless product $Q_{(\alpha,i)}$.
Therefore the factor $(kl)^{(i)}$ can be  replaced by
$ (Qk)^{(i)} (Ql)^{(i)} / (QQ)^{(i)}$. Then the factors
involved are expressed through ordinary products $Qk$ and $Ql$.

Now, an arbitrary integral $J(a_1,a_2,a_3,a_4,0)$
is reduced to integrals with $a_3=0$ with the help of
the following relation  obtained by
integration by parts
\be
(2D-2-2a_4-a_3)\mmm -(D-1-a_4)\mmmm + 4a_3\ppp = 0.
\label{eq:rela3}
\ee
In the present problem we only need the integrals $J$ with the unit
values of the first two arguments $a_1=a_2=1$.  In general it is also
easy to reduce $a_{1,2}$ to 1 using a similar recurrence relation.

The relation (\ref{eq:rela3})
can be used to express an arbitrary integral
$J(1,1,a_3,a_4,0)$ through integrals with $a_3=1$ and $a_3=0$.
The latter will be considered below. To calculate
$J(1,1,1,a_4,0)$ we use the symmetry with respect to
$l\leftrightarrow k$ and replace $2Ql$ in the numerator by
$Qk+Ql$, which decreases $a_4$.  In the resulting integral we use
partial fraction decomposition with respect to $2Ql$.  As a result we
get integrals with decreased $a_4$
and  products of one-loop integrals.
If $a_3$ becomes increased in this process we apply the relation
(\ref{eq:rela3}) again.


In the result of manipulations described above,
the only two-loop integrals we are left with
are of the form $J(1,1,0,a,0)\equiv J(a)$
\begin{eqnarray}
J(a) =
\int\!\!\int
\dr^Dk \dr^Dl
{1
\over
(k^2-m^2+i0)
(l^2-m^2+i0)
(2Qk+2Ql+i0)^{a}} \, .
\end{eqnarray}

In $J(a)$ we first
perform the $k_0$ and $l_0$ integrations using Cauchy theorem; the
angular
integrations are trivial since there are no products $kl$ in the
integrand.
The two remaining radial integrations lead to
\be
J(a)=
-{C^2 2^{1-2\vep}  \over (-4Mm)^a }
{\cos\left(\pi\vep+{\pi a\over 2}\right)  \over \cos(\pi\vep) }
B\left({3\over 2}-\vep, {a\over 2}-1+\vep\right)
B\left({a\over 2}-2+2\vep, -{a\over 2}+{3\over 2}-\vep \right)
\ee
with  $C = {2m^{2-2\vep} \pi^{{5\over 2}-\vep}
\over \Gamma\left({3\over 2}-\vep\right)}$.  This
formula is valid to all orders in $\ep$.

\section{The final result}

After adding the contributions (i), (ii), and (iii)
we obtain the following finite result
\be
G_2= {\pi^4\over M^2} \left\{ \pi^2\ln2 - {3\over 2}\zeta(3)
+\sum_{n=1}^{18}  \left[ a_n + \pi^2 b_n + \ln(x^2) c_n \right]  x^n
+{\cal O}(x^{19})\right\} \, .
\label{eq:G2}
\ee

\begin{table}[htb]
\begin{center}
\begin{tabular}{|r|l|l|}
\hline
$n$ & $a_n$ & $c_n$\\
\hline
\hline
1  &0 & $\pi$ \\
2  &$-1/2$ &0 \\
3  &2$\pi$/3 &  $\pi$/24\\
4  &$- 5/144$& $-1/12$ \\
5  &13$\pi$/60 & 3$\pi$/640\\
6  &107/5400 & $-1/15$\\
7 & 529$\pi$/6720 & 5$\pi$/7168\\
8 & 9073/470400 & $-73/1680$ \\
9 & 14887$\pi$/483840 & 35$\pi$/294912 \\
10 & 101923/7938000 & $- 17/630$ \\
11 & 715801$\pi$/56770560 & 63$\pi$/2883584 \\
12 & 3620783/461039040 & $-1391/83160$ \\
13 & 31515089$\pi$/5904138240 & 231$\pi$/54525952 \\
14 & 47658179/10100170080  & $-317/30030$\\
15 & 3278369671$\pi$/1416993177600 & 143$\pi$/167772160 \\
16 & 24219791/8480609280  & $- 19741/2882880$ \\
17 & 13114766971$\pi$/12847404810240 & 6435$\pi$/36507222016 \\
18 & 8503364437/4825201661280 &  $-21071/4594590$ \\
\hline
\end{tabular}
\end{center}
\caption{Coefficients of the expansion of the diagram $G_2$}
\end{table}
The coefficients $a_n$ and $c_n$ are given in Table 1.
The computing time for these first 18 powers of $x$ is of the order of
a few hours on a DEC ALPHA workstation.
For the coefficients of $\pi^2$ we find  $b_{2n+1}= 0 $,
$b_{2n}=-{1\over  n2^{n}}$.  If this is true for
all $n$ this part of the series can be summed up analytically:
$\pi^2\sum_{n=1}^\infty b_{2n} x^{2n} = \pi^2\ln(1-x^2/2)$.

Similarly to the one-loop result (\ref{eq:1loop}) we see in the
expansion (\ref{eq:G2}) non-analytical terms with odd
powers of $x$; at two-loop we also obtain 
logarithmic terms $\pi \left({m\over M}\right)^{2n+1}\ln
\left({m\over M}\right)$.

The expansion (\ref{eq:G2}) converges at least up to $m/M=1$.  We have
compared the values obtained with this formula with the numerical
program \cite{Baub}\footnote{We thank J.~Franzkowski for sending us an
 independent evaluation of $G_2$ at  $x=1$.}.  For $x<0.5$ the
accuracy is better than
$10^{-5}$.  Even at $x=1$ the error is only 0.7\%; at this point we
can use for comparison the recently found analytical value of the curly
bracket in eq.~(\ref{eq:G2}) at $x=1$, 
$-\zeta(3)+{2\over 3}\pi{\rm  Cl}_2(\pi/3)$ \cite{Jikia}.  
Much more important
than these numerical achievements is the possibility of extending the
expansion (\ref{eq:G2}) to any order in $x$. The methods
described in this paper can also be extended to other
cases of propagators and to vertex diagrams.  For example, the results
of ref.~\cite{b2c} for two-loop QCD corrections to $b\to c$
transitions at zero recoil can be extended to other kinematical
regions.  Work on this is in progress.

\section*{Acknowledgments}
We are grateful to P.A. Baikov, A.G. Grozin, and O.V. Tarasov for help
with checking the analytical results for the contribution (i).  This
research has been supported by the Russian Foundation for Basic
Research, project 96--01--00654, and by the grant BMBF 057KA92P.

\end{document}